# 5G from Space: An Overview of 3GPP Non-Terrestrial Networks

Xingqin Lin, Stefan Rommer, Sebastian Euler, Emre A. Yavuz, and Robert S. Karlsson

Ericsson

Contact: xingqin.lin@ericsson.com

*Abstract*— We provide an overview of the 3rd generation partnership project (3GPP) work on evolving the 5G wireless technology to support non-terrestrial satellite networks. Adapting 5G to support non-terrestrial networks entails a holistic design spanning across multiple areas from radio access network to services and system aspects to core and terminals. In this article, we describe the main topics of non-terrestrial networks, explain in detail the design aspects, and share various design rationales influencing standardization.

## I. INTRODUCTION

The 3rd generation partnership project (3GPP) completed the standardization of the first global 5th generation (5G) wireless technology in its Release 15 in mid-2018 [1]. The first evolution step of the 5G system was finalized in Release 16, and 3GPP is working on further evolution of the 5G system in Release 17. Enabling 5G system to support non-terrestrial networks (NTNs) has been one direction under exploration in 3GPP. The objective of this article is to provide an overview of the state of the art in 3GPP NTN work.

NTN has become an umbrella term for any network that involves non-terrestrial flying objects. The NTN family includes satellite communication networks, high altitude platform systems (HAPS), and air-to-ground networks, as illustrated in Figure 1.

Satellite communication networks utilize spaceborne platforms which include low Earth orbiting (LEO) satellites, medium Earth orbiting (MEO) satellites, and geosynchronous Earth orbiting (GEO) satellites. Over the past several years, the world has witnessed resurging interest in the broadband provisioned by LEO NTNs with large satellite constellations (e.g., Starlink, Kuiper, and OneWeb). To benefit from the economies of scale of the 5G ecosystem [2], the satellite industry has engaged in the 3GPP process to integrate satellite networks into the 5G ecosystem.

HAPS are airborne platforms which can include airplanes, balloons, and airships. In the 3GPP NTN work, the focus is on high altitude platform stations as International Mobile Telecommunications base stations, known as HIBS. A HIBS system provides mobile service in the same frequency bands used by terrestrial mobile networks.

Air-to-ground networks aim to provide in-flight connectivity for airplanes by utilizing ground stations which play a similar role as base stations (BSs) in terrestrial mobile networks. But the antennas of the ground stations in an air-to-ground network are up-tilted towards the sky, and the inter-site distances of the ground stations are much larger than that of terrestrial mobile networks.

So far, the focus of 3GPP NTN work has been on satellite communications networks, with implicit compatibility to support HIBS systems and air-to-ground networks. It is worth noticing that 3GPP has also been working on mobile enabled low-altitude unmanned aerial vehicles (UAVs, aka. drones), which can be considered as part of the NTN family in a wide sense. However, this line of work has been carried out in a separate track in 3GPP. Therefore, here we will mainly focus on the satellite communication networks and keep the treatment of other types of NTNs to a minimum.

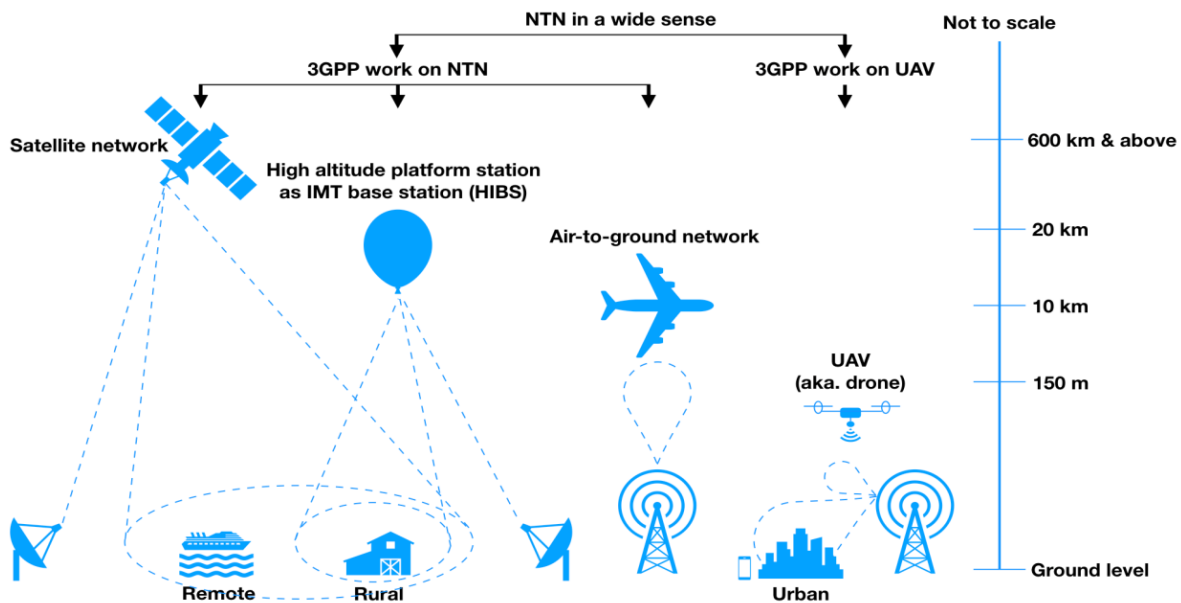

**Figure 1: Different types of non-terrestrial networks.**

5G New Radio (NR) based NTN has been the main focus in 3GPP [3]. NR was designed for forward compatibility, support for low latency, advanced antenna technologies, and spectrum flexibility including operation in low, mid, and high frequency bands. This provides a solid foundation for adapting NR to support NTNs.

There is growing interest in NTN based massive Internet of Things (IoT) use cases using narrowband IoT (NB-IoT) and Long-Term Evolution (LTE) for machine type communication (LTE-M). As a result, 3GPP is studying the feasibility of adapting NB-IoT and LTE-M to support NTN in its Release 17 [4].

An overview of the role of NTNs, application scenarios, and networking challenges is presented in [5]. The work in [6] discusses NTN challenges and opportunities and presents a case study on using millimeter wave frequencies to connect mobile terminals. A comprehensive survey on NTN is provided in [7], but the discussion on 3GPP NTN work therein stays at a high level. In contrast, the objective of this article is to offer a dedicated treatment of the 3GPP NTN work by delving into the detailed aspects and sharing the various design rationales influencing standardization. In particular, this paper discusses key NTN topics spanning across multiple areas from radio access network to services and system aspects to core and terminals.

## II. RADIO ACCESS NETWORKS FOR NR NTN

### A. Release-15 Study Item on NR NTN

3GPP work on NR NTN started in 2017, with a Rel-15 study focused on deployment scenarios and channel models. The study was documented in 3GPP TR 38.811 [8].

The first main objective of the study was to select a few reference deployment scenarios of NTN and agree on key parameters such as architecture, orbital altitude, frequency bands, etc. The key scenarios and models include
- Two frequency ranges, S-band and Ka-band
- GEO satellites, LEO satellites, as well as HAPS
- Earth-fixed beams (i.e., beams that are steered towards an area of earth as long as possible) and moving beams (i.e., beams that move over the Earth's surface following the motion of the satellite)
- Typical footprint sizes and minimum elevation angles for GEO, LEO, and HAPS deployments
- Two types of NTN terminals: handheld terminals and Very Small Aperture Terminals (VSAT) (equipped with parabolic antennas and typically mounted on buildings or vehicles)
- Antenna models for the satellite and HAPS antennas

The second main objective of the study was to develop NTN channel models based on the terrestrial 3GPP channel models.

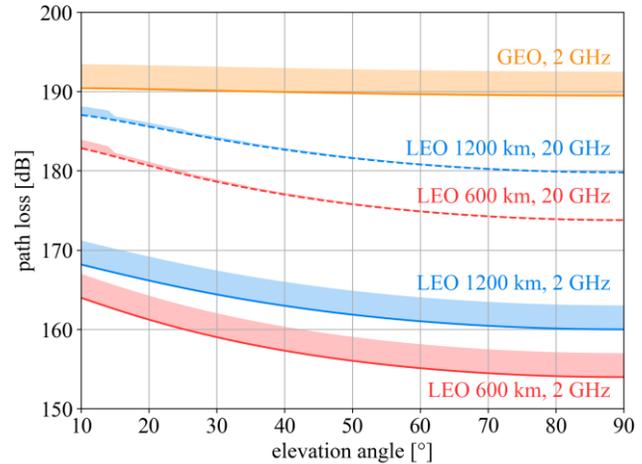

Figure 2: Path loss as a function of elevation angle

The channel models developed supports a range of deployment scenarios including urban, suburban, and rural.

Multipath is a typical phenomenon in terrestrial propagation environments. For NTN, the large distance to the satellite causes different paths to be almost parallel, and the angular spread is thus close to zero. The large-scale parameters (line-of-sight probability, angular spread, delay spread, etc.) are therefore different from the terrestrial case and depend on the elevation angle of the serving satellite.

Modeling of the path loss mainly relies on free-space path loss but adds components for clutter loss and shadow fading to account for the attenuation by surrounding buildings and objects. Values for clutter loss and shadow fading are tabulated for different elevation angles and for the two frequency ranges of S-band and Ka-band. The channel models also include parameters to account for absorption by atmospheric gases, as well as ionospheric and tropospheric scintillation losses. These losses may be of interest only for low elevation angles and/or in certain other conditions (e.g., at low latitudes, during periods with high solar activity, etc.). Figure 2 shows the path loss for various satellite orbits and carrier frequencies as a function of elevation angle, including atmospheric losses and assuming line-of-sight conditions. The shaded areas indicate the additional loss from scintillation, assuming a moderate scintillation strength.

The study developed two fast fading models. A more generic and frequency-selective model is based on the terrestrial model but adjusted to the satellite geometry with different values and correlations for delay and arrival angles. Similar to clutter loss and shadow fading, values are tabulated for different elevation angles and for the two frequency ranges. Alternatively, a simpler two-state model assuming flat fading can be used to study certain situations (e.g., low frequencies, large elevation angles, and near-line-of-sight).

In addition, the study developed clustered delay line and tapped delay line channel models for NTN link-level simulations.

### B. Release-16 Study Item on NR NTN

After completing the Rel-15 study on scenarios and channel models for NR to support NTN, 3GPP continued with a follow-up Rel-16 study on solutions for adapting NR to support NTN.

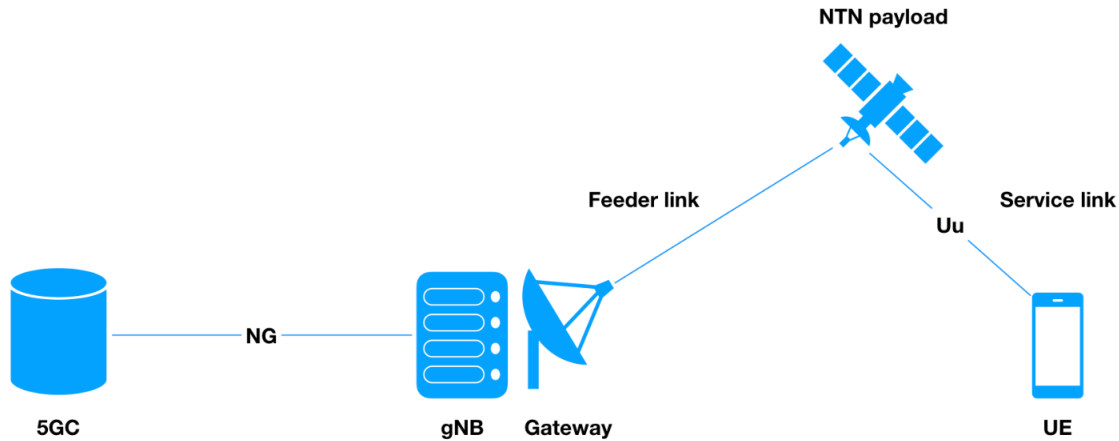

Figure 3: NTN architecture with transparent payload.

The main objective was to identify a minimum set of necessary features enabling NR support for NTN (especially for satellite communication networks). This included architecture, higher-layer protocols, and physical layer aspects. The outcome of the study is documented is documented in 3GPP TR 38.821 [9].

Next-generation RAN (NG-RAN) – the RAN for 5G – supports splitting a 5G base station (gNB) into a central unit (CU) and a distributed unit (DU). A multitude of NTN-based NG-RAN architecture options were explored. It was concluded that there are no showstoppers to support the identified architecture options.

The NR higher-layer protocol stack is divided into a user plane (UP), which manages data transmission, and a control plane (CP), which is responsible for signaling. For the UP, the main impact comes from the long propagation delays in NTN. Accordingly, the impact of long delays on medium access control (MAC), radio link control (RLC), packet data convergence protocol (PDCP), and service data adaptation protocol (SDAP) were studied. It was concluded that MAC enhancements would be needed for random access, discontinuous reception (DRX), scheduling request, and hybrid automatic repeat request (HARQ). It was recommended to focus on status reporting and sequence numbers on the RLC layer, and service data unit discard and sequence numbers on the PDCP layer. For the SDAP, it was found that it is not necessary to introduce any modification to support NTN.

For the CP, the focus of the study was on mobility management procedures, due to the movements of NTN platforms, especially LEO satellites. For idle mode, introducing NTN specific system information would be needed. Earth-fixed tracking area can be used to avoid frequent tracking area update. Defining additional assistance information for cell selection and reselection can be beneficial. For connected mode, handover enhancements were discussed to tackle the frequent handover due to the fast satellite movement.

From a physical layer perspective, extensive link level and system level evaluations were conducted in two nominal frequency bands: S band and Ka band. Based on the evaluation results, it was concluded that with appropriate satellite beam layouts, handheld user equipment (UE) can be served by LEO and GEO in S-band and that other UE with high transmit and receive antenna gains (e.g., very small aperture terminal (VSAT) and UE equipped with proper phased array antenna) can be served by LEO and GEO in both S-band and Ka-band. The study concluded that the Rel-15 and Rel-16 NR functionalities form a good basis for supporting NTN, despite issues due to long propagation delays, large Doppler shifts, and moving cells in NTN. Enhancements in the areas of timing relationships, uplink time and frequency synchronization, and HARQ were identified to be necessary.

### C. Release-17 Work Item on NR NTN

Based on the outcome of the Rel-16 study, 3GPP decided to start a work item on NTNs in NR Rel-17. The objective is to specify enhancements necessary for LEO and GEO based NTNs while also targeting implicit support for HAPS and air-to-ground networks. This involves the physical layer aspects, protocols, and architecture as well as the radio resource management, RF requirements, and frequency bands to be used. The focus is on transparent payload architecture with earth fixed tracking areas and frequency-division duplexing (FDD) systems where all UEs are assumed to have global navigation satellite system (GNSS) capabilities.

Figure 3 provides an illustration of the NTN architecture with transparent payload. The 5G core (5GC) network connects to a gNB using the NG interface. The gNB is located on the ground and connects to an NTN gateway that via the feeder link connects to the NTN payload (a network node embarked onboard a satellite or HAPS). The NTN payload connects to the UE via the service link using the Uu interface.

In terrestrial NR, the uplink timing is based on the downlink received timing and the propagation time is usually much smaller than a transmissions slot, while in NTNs the propagation time is much longer than a transmission slot.

A UE with GNSS capabilities can from its position and the NTN ephemeris calculate the relative speed between the UE and the satellite, as well as the round-trip time (RTT) between the UE and the satellite. From the relative speed the UE can calculate and apply a pre-compensation for the doppler frequency to ensure that its uplink signal is received at the satellite or at gNB on the desired frequency. The gNB provides the UE with a common timing advance (TA) that signals the RTT between the satellite and the gNB. The UE adds the RTT between the UE and the satellite to the common TA to get the full TA. The full TA is used as an offset between the received downlink timing and the uplink transmission timing at the UE, that is, if downlink slot n starts at time t1, then uplink slot n starts at time t1 minus the full TA. This enables the UE to send uplink transmissions with accurate received timing at the gNB for both random access and data transmissions in connected mode.

The transmissions in Rel-16 NR are based on up to 16 stop-and-wait HARQ processes for continuous transmissions. A HARQ process cannot be reused for a new transmission until the feedback for the previous transmission is received. With long RTTs and using stop-and-wait protocol, the transmissions will stall when all HARQ processes are waiting for feedback, which reduces communication throughput. To mitigate the stalling, the number of HARQ processes is extended to 32 which can cover some air-to-ground scenarios. The 32 HARQ processes are however not enough to cover the RTTs of LEO and GEO based NTNs. As further extension of the number of HARQ processes is deemed undesirable, schemes for reusing the same HARQ process before a full RTT has passed have to be employed to avoid stalling. When reusing a HARQ process for downlink transmissions before a RTT has passed, the HARQ feedback becomes unnecessary and thus is disabled. For uplink there is no HARQ feedback and the gNB can dynamically decide if a HARQ process shall be reused before a RTT has passed by sending grants for new data or grants for retransmissions, or wait until it has decoded an uplink transmission to decide to send a retransmission grant.

For HARQ processes with disabled feedback, to save energy the UE does not need to listen for retransmission assignments after a period. When HARQ is not used for retransmissions, the link adaptation may target a lower block error rate, but to achieve robustness a higher RLC retransmission rate, as well as more RLC status reporting, is expected.

To cover the long RTT in NTNs, some of the MAC and RLC timers are extended. As the satellites move, there is a need for the UE to (re)select a new satellite, which is based on the existing criteria and may include new criteria such as the timing when a satellite stops covering the area where the UE is located. Conditional handover is enhanced with a new condition based on UE location and the timing of satellite coverage of the UE location. Measurement procedures are enhanced with location-based triggering.

## III. SERVICES AND SYSTEM ASPECTS FOR NR NTN

### A. Services and Requirements

The 3GPP SA working group 1 (SA1) is responsible for the overall system requirements for 3GPP systems. In 2017, SA1 started a study to identify use cases and service requirements for using satellite access in 5G. The study, documented in TR 22.822 [11], identified use cases for satellites being used both as an access technology from a UE as well as a backhaul link between a terrestrial BS and a core network (CN). For the UE satellite access case, use cases include, e.g., use of satellite for broadcast service, to guarantee coverage for IoT devices, and to provide mission critical access in disaster situations. For the satellite backhaul scenarios, use cases include, e.g., fixed backhaul between a BS in a remote area and a CN, as well as backhaul between a moving BS deployed on a train and a CN.

The analysis of the use cases led to the formulation of service requirements that were then included in the overall service requirements specification for the 5G System (i.e., TS 22.261). The requirements cover both NTN RAN based satellite access for access and backhaul use cases, as well as the possibility to use satellite radio technology not developed by 3GPP.

In 2019, SA1 approved a new Rel-18 study to address aspects related to extra-territorial coverage of satellites and high-altitude systems. Current terrestrial 3GPP systems are typically deployed so that they provide coverage within a single country only, fulfilling the associated regulatory obligations for that specific country. Satellite-based radio systems may, however, cover multiple countries or cover international waters. This leads to new challenges for the 3GPP system. The study aims to provide guidelines on the fulfillment of relevant legal and regulatory requirements in such situations.

### B. Architecture

The 3GPP SA working group 2 (SA2) is responsible for overall system architecture of the 5G System. Based on the services requirements defined by SA1, SA2 identifies the main network functions, how these functions are linked to each other, and the information they exchange.

In 2018, SA2 initiated a study on architecture aspects for using satellite access in 5G. The main task of the study was to investigate the impacts of supporting satellite access and backhaul on the 5G system, with an aim to reuse the existing solutions defined for terrestrial 5G networks, including the 5GC network. The 5GC reference architecture, with the main network functions, is shown in Figure 4. One of the main parts of the study was to identify potential impact on the 5GC network due to NR NTN access, i.e., identify potential differences in functional behaviors and interfaces compared to terrestrial NR.

Mobility management is used to support mobility (e.g., handovers) and reachability (e.g., paging). Since NR NTN access is based on NR, the same mobility management features as defined for terrestrial can be reused, but some complications arise, e.g., in case of non-GEO satellites with moving cells. The 5GC assumes Earth-fixed tracking areas (TAs) and also that the cell identities (IDs) refer to specific geographical areas. Both TA identifiers and cell identifiers are used in 5GC and service layer as information about UE location. Assuming Earth-fixed TAs and that NTN RAN reports cell IDs that can be mapped to geographical areas ensures that 5GC and services layer can continue to use these identifiers as representation of a UE location, even if radio cells are moving across the Earth's surface.

As mentioned in Section III-A, the potentially wide multi-country coverage of satellite radio systems is a challenge when it comes to fulfilling regulatory requirements. The access and mobility management function (AMF) may therefore need to verify that the UE is located in an area (country) that the AMF

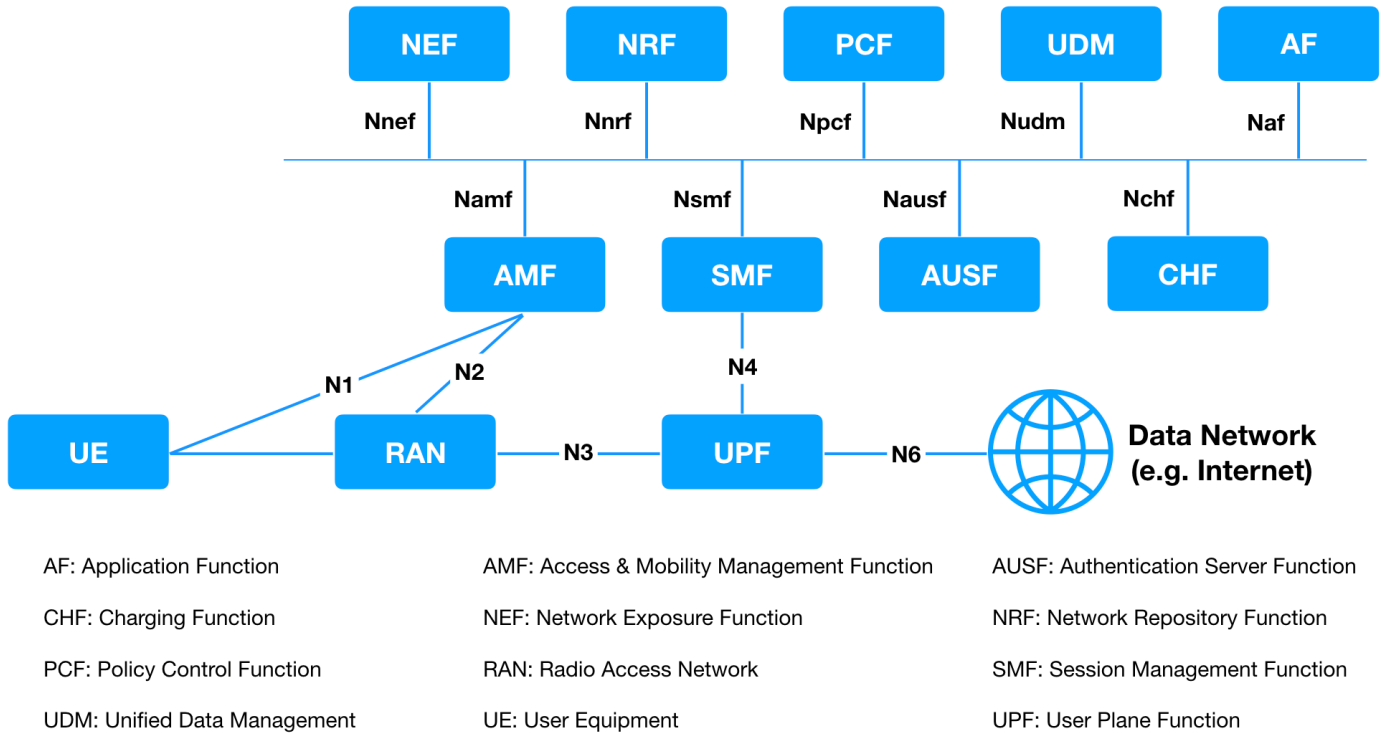

Figure 4: 5G system architecture (simplified).

is allowed to serve, as illustrated in Figure 5. The verification may be assisted by RAN.

The 5G quality of service (QoS) framework should be re-usable with small enhancements. In particular, when using a GEO satellite, the satellite connection may contribute significantly to the end-to-end delay which in many cases might be beyond what is today allowed by the standardized 5G QoS classes. Some adjustment to existing 5G QoS classes, or potentially definition of new 5G QoS classes, is likely needed. In addition, possibilities to inform the policy control function (PCF) and the application function (AF) about the use of satellite access or satellite backhaul may be beneficial, especially for long-delay GEO scenarios.

The 3GPP system today has significant possibilities for differentiated charging and policy control. One aspect that can be considered is the access technology used by the UE (e.g., whether the UE is using NR, or LTE, or WiFi). To enable this for the different NR NTN satellite types (LEO, MEO, GEO) and also allow differentiation with terrestrial NR, new radio access technology (RAT) type values will be introduced. This allows 5GC network functions for session management (SMF), policy control (PCF), charging (CHF) as well as the service layer (AF) to be aware of when the UE uses a satellite access (see Figure 4).

In summary, the conclusion of the study was that the 5GC, with small enhancements as described above, is well prepared to support NR NTN access as well as satellite backhaul. SA2 is currently working on producing the normative specifications for Rel-17, including satellite aspects.

*C. Telecom Management*

The 3GPP SA working group 5 (SA5) has the overall responsibility for management, orchestration, and charging for 3GPP systems. These include requirements, solutions, and protocol specific definitions.

In 2019, SA5 started a study on management and orchestration aspects with integrated satellite components in a 5G network. The main objective is to study business roles as well as service, network management, and orchestration of a 5G network with integrated satellite components. The scope includes both NTN RAN based satellite access, non-3GPP defined satellite access, as well as backhaul aspects. The aim is to reuse existing business model, management and orchestration of the current 5G network to minimize the impact.

The study outcome is documented in TR 28.808 [13] which includes use cases as well as potential requirements and solutions, e.g., for management and monitoring of gNB components and network slice management. Compared to terrestrial NR, the impacts mainly come from LEO/MEO scenarios where gNB components, such as gNB-DU, are located onboard satellite vehicles and would thus be moving relative to Earth. Other enhancements are needed due to the long delays that impact some of the monitoring functionality and key performance indicators.

The study concluded that the concepts of self-organizing networks (SON) for 5G would need to be enhanced to support mobile non-terrestrial gNBs. Another impact is the handling of performance measurements which make use of the HARQ process which may be unavailable when using satellite RAN with long delays. In addition, monitoring functions supporting the use of load balancing between different radio technologies is to be extended to cover load balancing between terrestrial RAN and non-terrestrial RAN. The work continues in Rel-17 and additional impact may be identified.

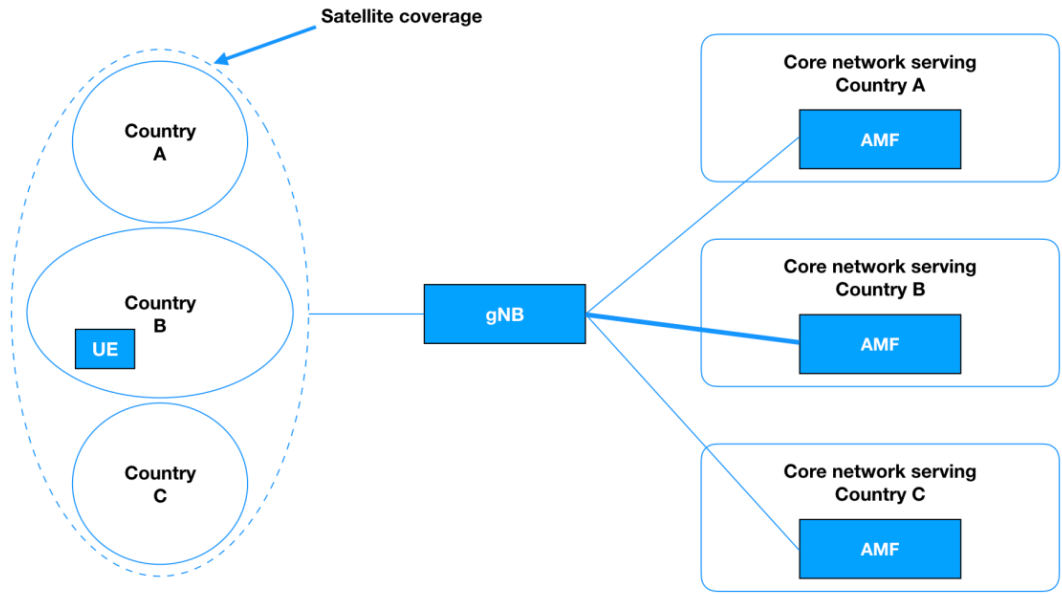

**Figure 5: Satellite access with satellite beams covering several countries and gNB connected to core networks serving the different countries.**

## IV. PROTOCOL AND NETWORK SELECTION ASPECTS FOR NR NTN

The core and terminals (CT) working groups in 3GPP define the core network protocols, as well as the protocol between the UE and the core network. The work on NR NTN has just started in these groups. We provide a brief overview of the ongoing work in this section.

The CT working group 1 (CT1) is responsible for the protocol between UE and the 5GC, as well as the UE's network selection. The main CT1 aspect when it comes to NR NTN is about how to handle public land mobile network (PLMN) selection. Legacy PLMN selection is defined primarily for terrestrial networks deployed within country borders. New aspects that arise for NR NTN include, e.g., how to handle PLMN selection in international areas (e.g., maritime environments), and how to handle global operators that use mobile country codes (MCC) that are not country specific. The regulatory requirements play a role here as it is assumed that the UE needs to select a PLMN with a core network in the country where the UE is located, even if it is the PLMN responsibility to fulfill the requirements and accept/reject UEs accordingly. This impacts PLMN selection, e.g., in cases where an NR NTN RAN covers multiple countries or is provided by a global operator. In addition to PLMN selection aspects, CT1 is also investigating whether non-access stratum protocol timers need to be extended to allow for the long delays that may arise with, e.g., GEO. The study is being documented in TR 24.821 [15].

The CT working groups 3 (CT3) and 4 (CT4) are responsible for the protocols between network functions in the core network. Satellite access and backhaul have small impacts on these interfaces, but as mentioned in Section III-B, a few enhancements are likely to be introduced:
- The possibility to signal the NTN NR satellite access category type to CN entities such as SMF, PCF and CHF. New values to be defined include "NR(LEO)", "NR(MEO)" and "NR(GEO)".
- The possibility to indicate to the PCF and to the AF that satellite backhaul with long packet delay is used. This allows the PCF to take the long delay into account when making policy and QoS decisions, and the AF to take the long delay into account on service layer.

The two aspects above allow, e.g., PCF to be informed about the use of satellite for both UE satellite access and satellite backhaul scenarios.

## V. IOT NTN

In Rel-13, 3GPP specified LTE-M and NB-IoT to support massive machine type communications (mMTC) to address the following design targets: low UE complexity and hence reduced device cost, long UE battery life to limit the need for charging and/or replacement, and coverage enhancements. 3GPP and International Telecommunication Union (ITU) have defined a set of 5G requirements for mMTC that can be met by both LTE-M and NB-IoT. For industries in areas with no cellular connectivity, connection may be provided via NTN to support mMTC to complement terrestrial deployments. In Rel-17, 3GPP has started a study to evaluate and confirm solutions to address the minimum necessary specifications for adapting LTE-M and NB-IoT to support NTN [4].

| Technical Specification Group | Release | Study Item / Work Item | Responsible Groups | Technical Report |
|---|---|---|---|---|
| **RAN (Radio Access Network)** | Rel-15 | RP-171450: Study on NR to support non-terrestrial networks<br>Objective: study channel model, deployment scenarios, and potential key impact areas. | RAN plenary, RAN1 | TR 38.811 [8] |
| | Rel-16 | RP-190710: Study on solutions for NR to support non-terrestrial networks<br>Objective: study a set of necessary features enabling NR support for NTN. | RAN1, RAN2, RAN3 | TR 38.821 [9] |
| | Rel-17 | RP-201256: Solutions for NR to support non-terrestrial networks<br>Objective: specify the enhancements identified for NR NTN with a focus on LEO and GEO and implicit compatibility to support high altitude platform station and air-to-ground scenarios. | RAN1, RAN2, RAN3, RAN4 | n/a |
| | Rel-17 | RP-193235: Study on NB-IoT/eMTC support for NTN<br>Objective: identify scenarios and study necessary changes to support NB-IoT and eMTC over satellite. | RAN1, RAN2 | TR 36.763 [10] |
| **SA (Service & System Aspects)** | Rel-16 | SP-170702: Study on using satellite access in 5G<br>Objective: identify use cases and the associated requirements. | SA1 | TR 22.822 [11] |
| | Rel-17 | SP-180326: Integration of satellite access in 5G<br>Objective: specify stage 1 requirements. | SA1 | n/a |
| | Rel-17 | SP-181253: Study on architecture aspects for using satellite access in 5G<br>Objective: identify key issues of satellite integration in 5G system architecture and provide solutions for direct satellite access and satellite backhaul. | SA2 | TR 23.737 [12] |
| | Rel-17 | SP-191335: Integration of satellite systems in the 5G architecture<br>Objective: produce normative specifications based on the conclusions identified in TR 23.737. | SA2 | n/a |
| | Rel-17 | SP-190138: Management and orchestration aspects with integrated satellite components in a 5G network<br>Objective: identify key issues associated with business roles, service and network management, and orchestration of a 5G network with integrated satellite component(s) and study the associated solutions. | SA5 | TR 28.808 [13] |
| | Rel-18 | SP-191042: Guidelines for extra-territorial 5G systems<br>Objective: study use cases of extra-territoriality, identify relevant features, technical aspects, and applicable types of regulations. | SA1 | TR 22.926 [14] |
| **CT (Core Network & Terminals)** | Rel-17 | CP-202244: CT aspects of 5GC architecture for satellite networks<br>Objective of study phase: study the issues related to PLMN selection and propose solutions.<br>Objective of normative phase: support the stage 2 requirements, and the requirements and solutions for PLMN selection for satellite access. | CT1, CT3, CT4 | TR 24.821 [15] |

**Table 1: A summary of 3GPP NTN work**

The first objective of the study is to identify the scenarios that are applicable to both LTE-M and NB-IoT, assuming that the bands used are in sub 6 GHz, satellite constellation orbits can be either LEO or GEO, payload is transparent, and nominal maximum device output power can be either 20 dBm or 23 dBm. The second objective is to study and recommend necessary changes to support LTE-M and NB-IoT over satellite for the identified scenarios in the first objective. In particular, aspects related to random access procedure and signals, HARQ operation, timers, idle and connected mode mobility, system information and tracking area are to be addressed. Both evolved packet core (EPC) and 5GC networks are assumed to be supported.

Both LTE-M and NB-IoT devices are assumed to have GNSS capability so that the UE can estimate and pre-compensate timing and frequency offset with sufficient accuracy for uplink transmission. The impact of GNSS position fix on UE power consumption will be studied. The impact of long RTT and the need to perform GNSS measurements on timing relationships are to be considered in the study. Disabling HARQ feedback is yet another aspect to study for understanding potential benefits and/or drawbacks. For idle mode mobility, cell selection/re-selection mechanisms of LTE-M and NB-IoT are used as baseline. Connected mode mobility is only supported for LTE-M. In addition to the legacy handover mechanism, potential enhancements for conditional handover are to be considered for both moving cell and fixed cell scenarios. For NB-IoT, legacy radio link failure mechanism is assumed to be the baseline for mobility in the study.

## VI. CONCLUSIONS

The continuous evolution of 5G technology aims to improve performance and addresses new use cases. The inherent flexibility of 5G technology provides a solid foundation for adapting it to support NTNs. NTNs, particularly satellite communications network, are complex systems, the design of which requires a holistic approach. To this end, multiple 3GPP working groups from RAN to SA to CT have been dedicating remarkable efforts to NTN design over multiple 3GPP releases. This article has provided a comprehensive overview of the state of the art in NTN work in 3GPP by discussing core topics of NTN, explaining in detail the design aspects, and sharing various design rationales influencing standardization. Table 1 provides a summary of the 3GPP NTN work.

Making 5G from space a reality requires efforts beyond standardization. As the first release of normative NTN standardization work is expected to be completed in 2022, it is important to develop early prototypes for validating key NTN design aspects and providing prompt feedback to standardization. It will be exciting to see how 5G will play a role in providing connectivity from space in years to come.

## ACKNOWLEDGEMENT

The authors thank Peter Bleckert, Mark Scott, and Mikael Wass for their valuable comments and suggestions.